\documentclass[journal,compsoc]{IEEEtran}

\usepackage{epsfig}
\usepackage{amssymb}
\usepackage{amsmath}
\usepackage{amsfonts}
\usepackage{amsthm}
\usepackage{mathtools}
\usepackage{alltt}
\usepackage{epstopdf}
\usepackage{algorithmic}
\usepackage[ruled]{algorithm}
\usepackage{multirow}
\usepackage{xcolor}
\usepackage{diagbox}
\usepackage{colortbl}
\usepackage{cite}
\graphicspath{ {image/} }
\usepackage{tabu}
\usepackage[justification=centering]{caption}

\newcommand{\MalFox}{{\tt MalFox}}
\begin{document}
	
	\title{MalFox: Camouflaged Adversarial Malware Example Generation Based on Conv-GANs Against Black-Box Detectors}
	
	\author{Fangtian Zhong~,
		Xiuzhen~Cheng,~\IEEEmembership{Fellow,~IEEE},
		Dongxiao~Yu,
		Bei~Gong,
		Shuaiwen~Song,
		Jiguo~Yu,~\IEEEmembership{Senior Member,~IEEE}
	\IEEEcompsocitemizethanks{
			\IEEEcompsocthanksitem Fangtian Zhong and Xiuzhen Cheng are with the Department of Computer Science, The George Washington University, Washington DC, USA.\protect\\
			E-mail: squareky{\_}zhong@gwu.edu, cheng@gwu.edu
		}
	\IEEEcompsocitemizethanks{
		\IEEEcompsocthanksitem Dongxiao Yu is with the School of Computer Science and Technology, Shandong University, Qingdao, Chian.\protect\\
		E-mail: dxyu@sdu.edu.cn
	}
	\IEEEcompsocitemizethanks{
		\IEEEcompsocthanksitem Bei Gong is with Faculty of Information Science, Beijing University of Technology, Beijing, China. \protect\\
		E-mail: gongbei@bjut.edu.cn
	}
	\IEEEcompsocitemizethanks{
		\IEEEcompsocthanksitem Shuaiwen Song is with the School of Computer Science, University of Sydney, Sydney, Australia.\protect\\
		E-mail: shuaiwen.song@sydney.edu.au
	}
	\IEEEcompsocitemizethanks{
		\IEEEcompsocthanksitem Jiguo Yu (Corresponding Author) is with the School of Computer Science and Technology, Qilu University of Technology (Shandong Academy of Sciences) and Shandong Computer Science Center (National Supercom- puter Center in Jinan), Jinan, China.\protect\\
		E-mail: jiguoyu@sina.com
	}
	}
	\markboth{this paper has been accepted by IEEE Transactions  on Computers in an upcoming issue for publication}%
	{\MakeLowercase{\textit{et al.}}}
	
	\IEEEtitleabstractindextext{
		\begin{abstract}
			Deep learning is a thriving field currently stuffed with many practical applications and active research topics. It allows computers to learn from experience and to understand the world in terms of a hierarchy of concepts, with each being defined through its relations to simpler concepts. Relying on the strong capabilities of deep learning, we propose a convolutional generative adversarial network-based (Conv-GAN) framework titled MalFox, targeting adversarial malware example generation against third-party black-box malware detectors. Motivated by the rival game between malware authors and malware detectors, MalFox adopts a confrontational approach to produce perturbation paths, with each formed by up to three methods (namely Obfusmal, Stealmal, and Hollowmal) to generate adversarial malware examples. To demonstrate the effectiveness of MalFox, we collect a large dataset consisting of both malware and benignware programs, and investigate the performance of MalFox in terms of accuracy, detection rate, and evasive rate of the generated adversarial malware examples. Our evaluation indicates that the accuracy can be as high as 99.0\% which significantly outperforms the other 12 well-known learning models. Furthermore, the detection rate is dramatically decreased by 56.8\% on average, and the average evasive rate is noticeably improved by up to 56.2\%.   
		\end{abstract}
		
		\begin{IEEEkeywords}
		 Adversarial Malware Examples, Deep Learning, Generative Adversarial Network, Malware
		\end{IEEEkeywords}
		
	}
	
	\maketitle
	
    \section{Introduction}
	\label{sec:introduction}
	
	\IEEEPARstart{D}{eep} learning has been broadly investigated and showed great merits in various applications \cite{AEOpportunities,intellfault}. Nevertheless, with its rapid development, security issues have raised serious alarms in recent years.
	%
	Various attacks on AI algorithms and models have emerged. At the training process, poisoning attacks can undermine the original probability distribution of the training data by injecting malicious examples to lower the predictive precision of the model. At the test or inference process, evasion attacks can confuse a target system by feeding carefully designed input examples without changing the model \cite{adnet}. Szegedy \emph{et al.} formally proposed the concept of adversarial examples \cite{adverexample}, which can be constructed 
	by applying small but worst-case perturbations. These examples can make the model outputting incorrect answers with high confidence. More seriously, even though different machine learning systems have different architectures and training datasets, the same set of adversarial examples demonstrate great transferable abilities to attack related models \cite{adverLanguage}. As a result, a variety of adversarial examples used in different areas are thrived \cite{adverLanguage,adverText,adverImage,adverSpeech,adverAutoDriv}, which are summarized as follows.
	
	In the study of natural language processing, Zhang \emph{et al.} 
	presented an adversarial example generator by referring to the Metropolis-Hastings attack algorithm to develop fluent adversarial examples for natural languages, and the examples successfully mislead the bi-LSTM and BiDAF models on IMDB and SNLI datasets \cite{adverLanguage}. In text classification, Ebrahimi \emph{et al.} proposed an efficient method to produce white-box adversarial examples, greatly decreasing the accuracy of a character-level neural classifier \cite{adverText}. In image processing, Qiu \emph{et al.} proposed SemanticAdv to create semantically realistic adversarial examples via attribute-conditioned image editing, which achieves high targeted attack success rates under both white-box and black-box settings \cite{adverImage}. In automatic speech recognition, Alzantot \emph{et al.} presented a demonstration of adversarial attacks against the speech classification model with $87\%$ success rate by adding small background noises, without knowing the underlying model parameter and architecture \cite{adverSpeech}. In autonomous driving, Kong \emph{et al.} presented PhysGAN to generate realistic and physical-world-resilient adversarial examples for attacking common autonomous driving scenarios \cite{adverAutoDriv}. 
	
	Nevertheless, there have been no successful attacks against the collection of practical antivirus products and online scan engines, \emph{e.g.} VirusTotal \cite{virustotal}. In this paper, we aim at generating adversarial examples that can bypass the detection by VirusTotal. One of the major challenges herein is to edit malware files. Additionally, for a practical application, any generated adversarial example shall be functionally indistinguishable from its original file.
	
	To tackle these challenges, we propose MalFox, an intelligent adversarial malware example generation framework based on convolutional generative adversarial networks (Conv-GANs). MalFox consists of 5 major components: PE Parser, Generator, PE Editor, Detector, and Discriminator. PE Parser extracts features from malware and benignware and transforms them into vectors serving as the input to Generator and Discriminator. Generator produces perturbation paths based on the malware features and Gaussian noises to make Discriminator misclassify malware as benign, while Discriminator is trained to distinguish benignware and malware precisely. PE Editor is used to generate adversarial malware examples by following the perturbation paths and makes the change functionally indistinguishable from the original malware. Detector is used to label adversarial malware examples such that the labels can be exploited by Discriminator to provide feedback to Generator. We adopt VirusTotal as the MalFox Detector, as it provides a collection of practical antivirus products and online scan engines that are well-received by security professionals and public users. Our multi-fold contributions can be summarized as follows. 
	
	First, in this paper, we propose an intelligent Conv-GAN-based framework, namely MalFox, which can generate functionally indistinguishable adversarial malware examples by successfully editing malware files following customized perturbation paths. Our framework is flexible in that it can accept any number and type of perturbation methods, thus enlarging the search space for more effective perturbation paths that can improve the abilities of malware to evade detection. In our implementation, we develop three novel framework perturbation methods whose combinations help produce adversarial malware examples that can mislead third-party black-box malware detectors with high probability under realistic environments.

	Second, distinctive from other GAN-based adversarial malware example generation methods that attack self-developed scan engines based on machine learning models, MalFox intends to attack a collection of antivirus products and online scan engines provided by third parties, which renders it a more practical and powerful attack framework. We make use of VirusTotal, which is adopted by realworld applications. On the other hand, existing GAN-based methods cannot generate practical adversarial malware examples since they just simply inject benignware's function names as features into malware feature vectors without actually implementing them in the corresponding malware files, which implies that they cannot attack antivirus products and online scan engines in practice. To our best knowledge, MalFox is the first framework that can launch adversarial attacks targeting practical malware detectors from third parties, under black-box settings. 
	
	Third, we successfully conduct attacks on 82 malware detectors and search engines in VirusTotal without knowing their underlying implementation details. The results demonstrate that MalFox possesses outstanding performance in its ability of evading from detection. More specifically, MalFox can produce adversarial malware examples that can successfully confuse Discriminator with an accuracy of 99.0\%, and greatly lower the detection rate up to 56.8\% in VirusTotal. Besides, the average evasive rate reaches 56.2\%, which indicates that the generated adversarial malware examples may largely escape from detection as a regular user typically installs only one or a few antivirus products (much less than 82).
	
	
	
	
	The rest of the paper is organized as follows. Section~\ref{sec:background} provides an overview on popular malware evasion techniques. Section~\ref{sec:design} presents the design of MalFox. Section~\ref{sec:implementation} details the implementations of MalFox components. Section~\ref{sec:evaluation} evaluates the performance of MalFox, and Section~\ref{sec:conclusion} concludes the paper with a  future research discussion.

	\section{Background and Related Work}
	\label{sec:background}
	Traditional antivirus products and online scan engines were built by running virtual machines (VMs) over one of the available virtualization platforms such as VMware, VirtualBox, KVM, and Xen, which leaves huge gaps for malware to manipulate. \emph{Evasion} is an action in which malicious payloads exploit these gaps to avoid detection. 
	As malware detectors become more and more sophisticated and keep on evolving to defeat emerging evasion techniques, malware authors have to  change their tactics continuously in order to remain one step ahead. 
	In this section, we provide an overview on existing popular malware evasion methods, classifying them as either artificial intelligence (AI) based or non-AI based, as AI provides powerful tools for smart evasion solutions in recent years.
	
	\subsection{Non-AI Based Evasion Techniques}
	
	The most prevalent non-AI based evasion techniques include \emph{delaying execution}, \emph{fingerprinting}, \emph{obfuscation}, and \emph{multi-stage evasion}, which are summarized as follows.
	
	\textbf{Delaying Execution.} Due to the fact that it is computationally costly to run a complete detection environment, detectors usually only accurately visualize the behaviors of malware within a short period of time. This vulnerability can be exploited by malware to delay the normal execution of its malicious process.  
	Evasion techniques based on delaying execution include manipulating the delay application programing interfaces (NtDelayExecution, CreateWaitTableTimer, SetTimer, \emph{etc.}), sleep patching, and time bombs. Many malware programs exploit these techniques to successfully bypass anti-virus products. For example, in 2011, the Khelios botnet, which is capable of sending roughly 4 billion spam messages a day, called the NtDelayExecution() API with a 10-minute extended sleep period to evade detection \cite{Khelios}. Black POS malware, one of the most pervasive types of point-of-sale malware observed in-the-wild in 2013, exhibited time bomb evasion by executing during certain periods while remaining dormant at the rest of the time and successfully disguised an installed service of a known anti-virus software (KrebsOnSecurity). Consequently, customers' debit and credit card information was exposed to the public \cite{BlaPOS}. In 2012, the Trojan UpClicker employed the SetWindowsHookEX() API function to hide its malicious activity. By sending OEH as the parameter to the function, the malicious code was activated only when the left mouse button was clicked and released. Because most file-based detectors do not mimic human interactions, this malware remained dormant during analysis and successfully evaded detection \cite{Khelios}.
	
	\textbf{Fingerprinting.} Fingerprinting is a technique to detect the signs that attest to the presence of an analysis environment or debuggers of malware detectors. Major fingerprinting techniques employed by malware include analyzing Process Environment Block (PEB), searching for breakpoints, and probing for system artifacts. PEB is a data structure that exists per-process in Windows, and the fields in PEB contain information that can be retrieved by malware to detect the presence of a debugger \cite{PEB}. The most obvious one is a field named BeingDebugged. 
	Anti-debugging tactics relying on PEB constitute the critical part of evasion techniques observed in malware \cite{MALSOF}. Besides, to analyze malware, detectors often set breakpoints during the execution of malware and save the breakpoint address in CPU DR registers. The behavior of detectors can be spotted by the malware through a self-scan or integrity check. For example, the malware CIH uses GetThreadContext to check the CPU register and erases the breakpoint information to avoid analysis \cite{CIH}. Moreover, from installation to configuration and execution, detectors often leave traces behind in different levels of the OS, \emph{e.g.} in the file system, registry, process name, \emph{etc.} Hence, malware can simply look for these traces and wait for opportunities to activate codes.

	
	\textbf{Obfuscation.} Obfuscation concerns the practice of deliberately degrading the quality of information in some ways, to protect the privacy of the individual to whom that information refers. This technique converts a program into a form functionally equal to the original one but may make the program hard to be understood. At the early stage, obfuscation was used to protect the intellectual property of software developers by hiding the codes from the public. However, it has been widely used by malware authors to elude detection in recent years. Sharif \emph{et al.} presented a malware obfuscation technique that automatically conceals the trigger-based behavior of malware from input-oblivious malware analyzers \cite{codeobfus}. Their technique transforms specific branch conditions that rely on inputs by incorporating one-way hash functions such that it is hard to identify the values of the variables for which the conditions are satisfied. The conditional code is identified and encrypted with a key that is derived from the value satisfying the condition \cite{codeobfus}. Schrittwieser \emph{et al.} introduced a novel approach for code obfuscation called covert computation and raised the bar for semantic-aware code analysis \cite{covertcomputa}. Because current malware detection approaches ignore the fundamental knowledge of the underlying hardware, and the instructions cannot completely express the state of a microprocessor, this vulnerability was exploited to hide the implementation of certain specific functionality of malware \cite{covertcomputa}.
	
	\textbf{Multi-stage Evasion.} Detectors via data signature scanning and behavior signature scanning assume that a holistic malware is present in the analysis environment and is examined to trigger an alert. However, malware with multi-stage delivery can bypass such an analysis, as it includes a set of sequential stages, wherein each stage does not trigger the alert. For example, Ispoglou \emph{et al.} proposed malWash, a dynamic diversification engine that hides the behaviors of malware by distributing the malware's execution across many processes. Target malware is decomposed into small components that are then executed in the context of other processes. A master as a coordinator connects these components and transfers the execution flow among the different processes \cite{malwash}. Pavithran \emph{et al.} proposed D-TIME, a new distributed threadless independent malware execution framework to evade runtime detection. They chopped malware into small chunks of instructions and executed one chunk at a time in the context of an infected thread. The coordination between chunk executions is organized by shared memory with asynchronous operations \cite{DTIME}.
	
	The evasion techniques mentioned above play their roles via different tricks. Specifically, delaying execution, fingerprinting, and obfuscation mainly evade detection by advanced techniques outperforming detectors, while multi-stage evasion depends on other processes under the same runtime environment.

	\subsection{AI-Based Evasion Techniques}
	Artificial Intelligence (AI) essentially comprises algorithms capable of processing and learning from vast amounts of data and then making decisions autonomously. With the advancement of AI, detectors using AI techniques are common. Hence, malware evasion powered by AI appears in our vision recently. Such techniques can be classified as \emph{biology-inspired} and \emph{deep neural network-based (DNN-based)}.
	
	\textbf{Biology-Inspired Approaches.} In \cite{AIMED}, Castro \emph{et al.} used genetic programming that makes malware samples evolve to attack four static learning classifiers (namely Gradient Boosted Decision Tree, Sophos, ESET, and Kaspersky) until they are no longer able to precisely detect. Each malware is sent to a manipulation box to inject byte-level perturbations, and then different malware mutations are produced. Those malware mutations are sent to classifiers to receive  scores according to their evasive capability. The two with the highest scores are considered as the most valid evasion and continue to generate next generation which would inherit their evasive abilities. The children repeates the above process and ends up with the strongest malware mutation. In \cite{swarmvirus}, Zelinka \emph{et al.} put forward a scheme implemented by swarm intelligence-based algorithms that can smoothly evade machine learning-based detectors. This scheme emulates the behavior of the biological swarm systems that don't require a central communication point. The communications between members in the swarm system can be realized by having a collective memory via which every individual can share knowledge as well as "learn from it". Therefore detectors that exploit patterns extracted from a central communication point can not detect the abnormality.  
	
	\textbf{DNN-Based Approaches.} Fang \emph{et al.} proposed a framework named DQEAF, which uses reinforcement learning to evade machine learning-based detections \cite{reinforce}. This framework is composed of an attacking model and an AI agent. The attacking model is a series of functionality-preserving actions to alter malware samples. The agent inspects malware samples and chooses a sequence of actions to deliberately modify the samples for evasion. In the end, the framework can determine an optimal sequence of actions to evade detection. In \cite{andriodmalware}, Grosse \emph{et al.} performed adversarial crafting attacks on neural networks for malware classification. They generated adversarial  examples by adding entries to the AndroidManifest.xml file in malware. This scheme ensures that the addition of perturbations does not affect the utility of the malware.

	Papernot \emph{et al.} introduced the first demonstration that black-box attacks against detectors implemented by DNN classifiers are practical for real-world adversaries with no knowledge about the classifiers \cite{PraBlack}. 
	Their attack strategy is to train a local substitute DNN with a synthetic dataset: the inputs are synthetic and generated by the adversary while the outputs are labels assigned by the target DNN and observed by the adversary. Specifically,  some features extracted from the malware are replaced to generate an adversarial example serving as the synthetic input to mislead the local substitute DNN classifiers.
	Because the local and target models have similar decision boundaries, the attacks to the target DNN classifiers are successful if the malware evades the detection of the local DNN classifiers. Hu \emph{et al.} proposed a generative adversarial network (GAN) based algorithm named MalGAN to generate adversarial malware examples, which can bypass black-box machine learning-based detectors. MalGAN consists of a substitute detector, a generator, and a black-box malware detector. The substitute detector is trained to fit the black-box detector, while the generator is used to minimize the probability of malware being detected by the substitute detector via adding new features to the malware \cite{malgan}.
	
	The AI-based malware evasion techniques mentioned above build models based on a varity of views. Genetic programming and swarm-based intelligence algorithms emulate the behaviors of animals, while other approaches are on the basis of deep neural networks. In this paper, we present MalFox, an intelligent adversarial malware example generation framework based on Conv-GANs. Compared to the above-mentioned techniques, MalFox is able to generate practical functionality-preserving adversarial malware examples by making use of the power of classical and AI methods, and conduct practical attacks against a collection of antivirus products and online scan engines provided by third parties even without knowing the underlying implementation details. As a comparison, MalFox attacks the realworld malware detectors from third parties that exploit AI-based as well as other unknown techniques, while the models provided in \cite{reinforce,andriodmalware,PraBlack,malgan}  heavily rely on the powerful recognition capability of self-developed machine learning-based detectors. Moreover, \cite{andriodmalware, PraBlack,malgan} actually do not generate adversarial malware examples that can be adopted in practice. 
	
	
	
\section{MalFox Design}
	\label{sec:design}
	
	\begin{figure}[!htb]
		\centering
		\includegraphics[width=80mm,scale=0.25]{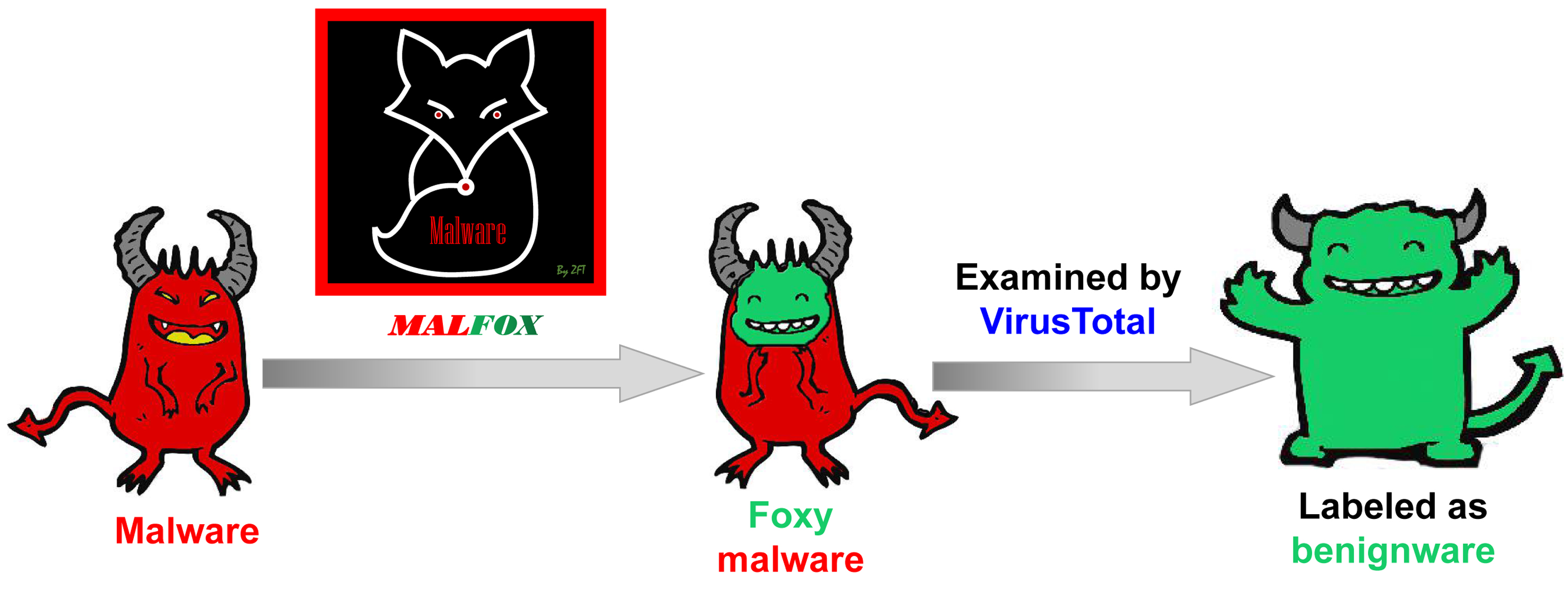}
		\caption{The MalFox Framework}
		\label{fig:fox}
	\end{figure}
	\begin{figure}[!htb]
		\centering
		\includegraphics[width=90mm,scale=0.25]{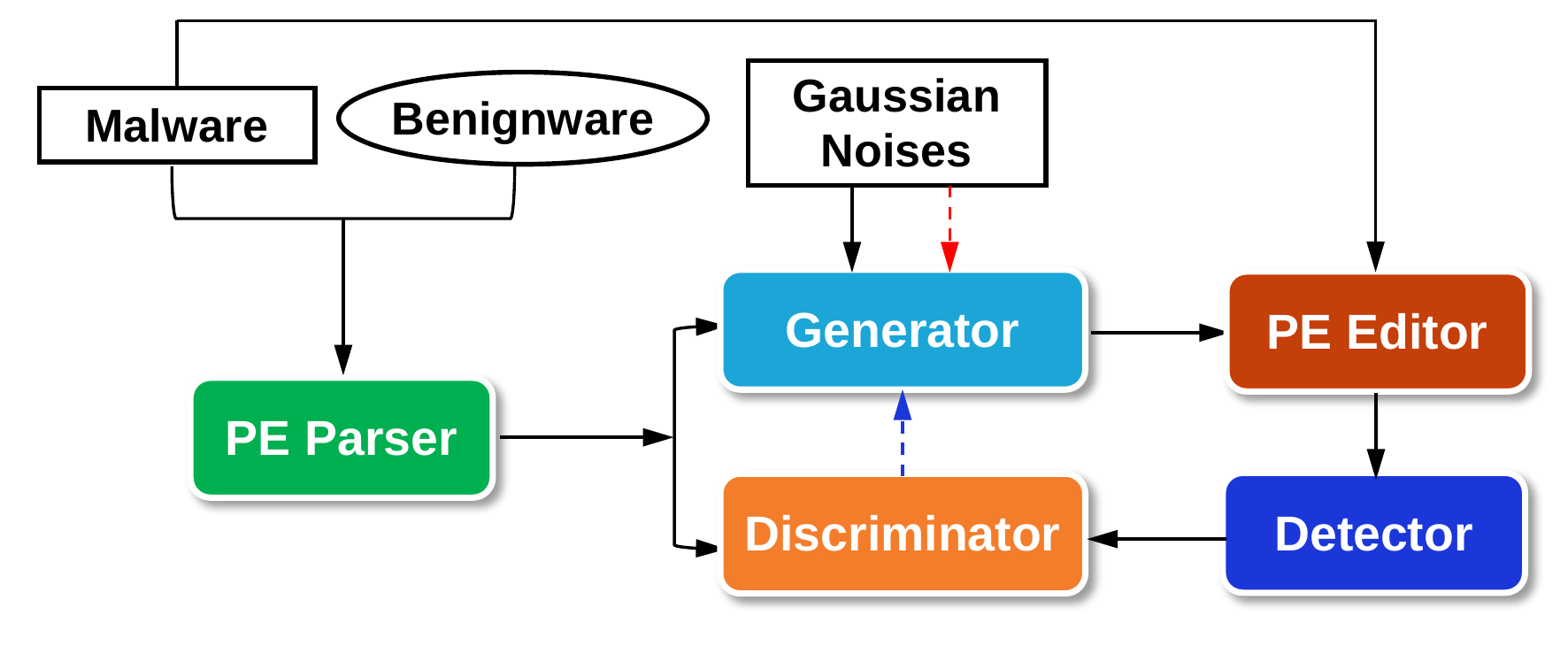}
		\caption{Architecture of MalFox}
		\label{fig:malfox}
	\end{figure}
	
	The objective of MalFox is to transform malware programs into foxy ones that are expected to be examined as benign by third party detectors such as VirusTotal (see Fig. \ref{fig:fox}). Fig.~\ref{fig:malfox} demonstrates the architecture of MalFox, which consists of 5 components: \emph{PE Parser}, \emph{Generator}, \emph{PE Editor}, \emph{Detector}, and \emph{Discriminator}. 

	PE Parser is developed to parse malware and benignware programs, and retrieve the dynamic link libraries (DLLs) and system functions that are called by each program under runtime environment. The names of DLLs and system functions are used as features by which MalFox can specify particular malware or benignware. When the names of all DLLs and system functions are extracted from the available malware and benignware under study, a feature set containing unique features is defined. We represent all features in the set as a vector; then PE Parser can denote each malware and benignware by a binary feature vector  with each element corresponding to the presence (`1') or absence (`0') of a certain feature. 
	
	Generator and Discriminator are components of the generative adversarial network. They are used to help generate an adversarial malware example for the input malware in MalFox. Specifically, for each malware program, Generator produces a perturbation path that is used by PE Editor to construct an adversarial malware example, which is expected to confuse Discriminator and finally escape detection by malware detectors. Discriminator, on the other hand, attempts to precisely distinguish malware and benignware. When Discriminator correctly identifies an adversarial malware example during the training process, Generator has to repeat its process to generate a new perturbation path for the malware; the process  stops when Discriminator fails to identify the adversarial malware example, at which time the PE Editor outputs the desired adversarial malware example. Therefore one can see that the stronger the Discriminator, the more powerful the Generator (and the PE Editor). 
	
	PE Editor manages the procedure of generating adversarial malware examples by following the perturbation paths produced by Generator. The implementation of PE Editor is the most challenging one in MalFox. We propose three novel framework methods, namely \emph{Obfusmal}, \emph{Stealmal}, and \emph{Hollowmal} in this paper, with each framework method supporting various flexible implementations and each perturbation path consisting of zero or more of the instantiated methods. 
	Obfusmal encrypts the code segment of a malware file and attaches a DLL named $Shell.dll$ at its end. $Shell.dll$ is responsible for decrypting the code segment and recovering the normal execution of malware. 
	Stealmal encrypts the entire malware and attaches it to the end of a program named $Shell.exe$. The duty of $Shell.exe$ is to decrypt the malware, create a suspended process, obtain the process space, copy the malware into the space, change the context of the process to the entry point of the malware, and resume the process. 
	Hollowmal encrypts the malware and attaches the encrypted malware into the end of a benignware program. Then, a DLL named $Hollow.dll$, which embraces similar functionality as $Shell.exe$, is added into the end of the modified benignware. All these three methods would not affect the normal execution of the malware. 
	Note that we say Obfusmal, Stealmal and Hollowmal are framework methods because the implementations of the corresponding $Shell.dll$, $Shell.exe$ and $Hollow.dll$ are not specifically defined, which implies that the space of the instantiated perturbation methods is very large, significantly enhancing the evasiveness of the produced adversarial malware examples. Particularly, $Hollow.dll$ employs a benignware program, which can be any benign program, further enlarging the search space of the perturbation methods.
	
	In our study, we employ VirusTotal as the MalFox Detector. VirusTotal is a webiste launched in June 2004, and deveoped by the Spanish security company Hispasec Sistemas, a subsidiary of Google Inc., which aggregates many antivirus products and online scan engines to check for malware. The virtualization solution used by VirusTotal is the Cuckoo sandbox. Users can upload files up to 550 MB to the website and receive the detective outcome from the antivirus products and online scan engines. Currently, VirusTotal handles approxiamtely one million submissions each day. The results of each submission are then shared with the entire community of antivirus vendors who lend their tools to the VirusTotal service, which in return allows vendors to benefit by adding into their products the malware signatures of new variants that their tools have missed but a preponderance of other tools have flagged as malicious \cite{statisvirus}. Popular tools such as McAfee, F-Secure, Tencent, 360, and Microsoft in VirusTotal, have been widely adopted on laptop and mobile devices. In this study, we use VirusTotal to ensure the reliability of our datasets, provide labels for Discriminator, and validate the performance of MalFox.
	
	
\begin{algorithm}[!htb]
	\caption{MalFox Training Procedure}
	\label{malfox}
	\begin{algorithmic}[1]
		\STATE{Convert each malware and benignware program in the training dataset into a binary feature vector by PE Parser;} 
		\WHILE{not converging}
		\STATE{Sample a minibatch of malware feature vectors and three-dimensional Gaussian noises, combine each malware feature vector with a noise sample, and input the results to Generator;}
		\STATE{Generator generates perturbation paths and inputs them to PE Editor;}
		\STATE{PE Editor produces adversarial malware examples following the perturbation paths;}
		\STATE{Sample a minibatch of benignware feature vectors;}
	    \STATE{Update Discriminator's parameters with the adversarial malware examples and benignware  programs by descending along the gradient of $L_D$ (Eq.\eqref{discriminator});}
		\STATE{Sample three-dimensional Gaussian noises, combine each with a malware feature vector in the minibatch, and input the results to Generator;}
		\STATE{Generator generates perturbation paths and inputs them to PE Editor;}
		\STATE{PE Editor produces adversarial malware examples following the perturbation paths;}
		\STATE{Detector labels the adversarial malware examples;}
		\STATE{Update Generator's parameters with the newly generated adversarial malware examples by descending along the gradient of $L_G$ (see Eq.\eqref{generator})}
		\ENDWHILE
	\end{algorithmic}
\end{algorithm}

	The procedure for constructing $\MalFox$ is illustrated by Algorithm~\ref{malfox}. Each benignware and malware program in the training dataset is first sent to PE Parser to get a binary malware feature vector (Line 1). A while loop follows to train $\MalFox$ iteratively (Lines 2-13) until Generator and Discriminator are stable (their model parameters do not change much from iteration to iteration). At each iteration, a minibatch of malware feature vectors are combined with Gaussian noises (one noise sample for each malware to get a noised malware program) as inputs to Generator (Line 3). Generator produces a perturbation path for each noised malware (Line 4), which is employed by PE Editor to edit the malware for producing the corresponding adversarial malware example (Line 5). Then the adversarial malware examples and a minibatch of benighware programs are used to train Discriminator (Line 6-7). 
	Next, a new set of three-dimensional Gaussian noises is produced and combined with the malware feature vectors selected in line 3 as inputs to Generator (Line 8), which constructs a new perturbation path for each noised malware program (Line 9). PE Editor follows the perturbation paths to generate adversarial malware examples (Line 10). Detector labels the adversarial malware examples (Line 11), and the labeled adversarial malware examples are used to trained Generator (Line 12).

Producing an adversarial malware example by MalFox for a given malware program is simple. The malware program is first transformed into a binary feature vector by PE parser, then combined with a random 3-dimensional noise to get the input to Generator. Generator outputs a perturbation path for the malware and the PE Editor modifies the original malware according to the perturbation path to produce an adversarial malware example.

	

	\section{MalFox Implementation}
	\label{sec:implementation}
	
	In this section, we detail our implementations of the four components of MalFox, namely PE Parser, Generator, PE Editor, and Discriminator.
	
	\subsection{PE Parser}\label{sec:PE:Parser}
	As mentioned in Section~\ref{sec:design}, the objective of PE Parser is to obtain features from benignware or malware programs and transform them into binary feature vectors. Since each malware or benignware was developed to satisfy certain functionality by employing a specific set of DLLs and calling a number of system functions, the names of these DLLs and system functions can be used as features to differentiate different benignware or malware. For example, to realize the special memory leakage check module, Avant Browser makes use of kernel32.dll, ADVAPI32.dll, COMCTL32.dll, CompareFileTime, GetShortPathNameA, GetFullPathNameA, MoveFileA, and SetCurrentDirectoryA, just to name a few.
	
	\begin{figure}[!htb]
		\centering
		\includegraphics[width=105mm,height=50mm]{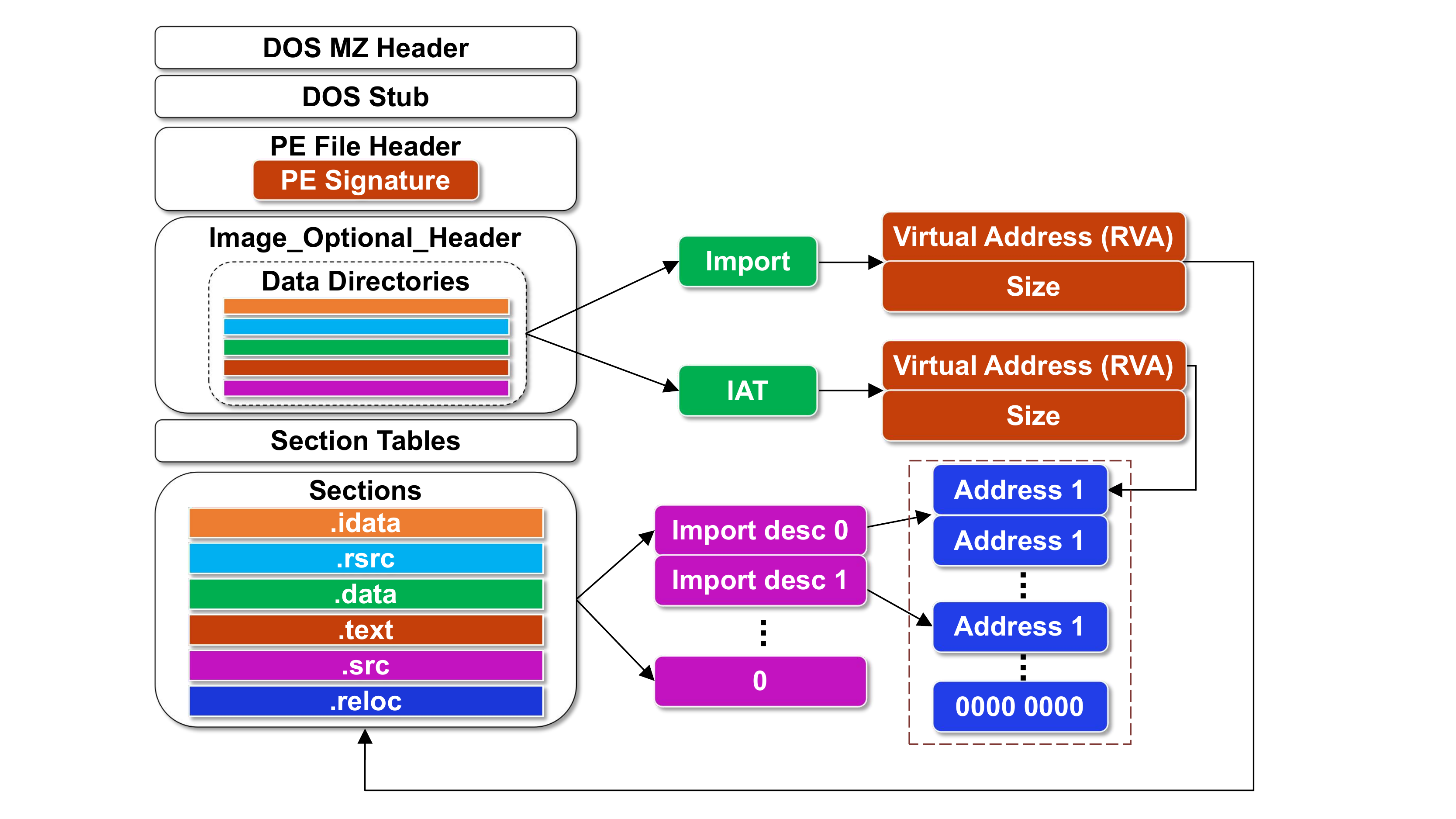}
		\caption{Structure of PE}
		\label{fig:PE}
	\end{figure}
	

	We collected a dataset with benignware and malware that follows the portable executable format (PE) used in Windows Operating Systems. To retrieve the names of the DLLs and system functions from the dataset, we need to take a close look at the structure of the PE as shown in Fig. \ref{fig:PE}. One can see that the PE has 6 components: $DOS\ MZ\ Header$, $DOS\ Stub$, $PE\ File\ Header$, $Image\_Optional\_Header$, $Section\ Tables$, and $Sections$, with each occupying certain number of bytes and including information related to different aspects of the PE. More specifically, DOS\ MZ\ Header has 64 bytes and  contains 19 fields, including the signature field that tells whether the program follows the PE and the address field that directs to PE\ File\ Header; DOS\ Stub has 224 bytes and intends to print linkers' information, which implies that the runtime environment is MS\_DOS; PE\  File\ Header has 64 bytes and specifies the machine type, the number of sections, and the size of Image\_Optional\_Header; Image\_Optional\_Header contains 224 bytes for 32-bit programs or 240 bytes for 64-bit programs and includes additional information such as ImageBase, SectionAlignment, FileAlignment, SizeOfImage,  SizeOfHeaders, and DataDirectories, which are required by the linker and loader in Windows; Section Tables is an array possessing the IMAGE\_SECTION\_HEADER structure 
with each entry containing information about one section in the Sections component; Sections contains the contents of the file, e.g. code, data, resources, and other executable information.
	
	To obtain the features of the benignware and malware programs, PE Parser needs to parse DataDirectories, which indicates where to find important components of the executable information in a file. DataDirectories has 16 tables, and each table references a data directory with an 8-byte entry, among which the Import (\texttt{Import}) and Import Address Table (\texttt{IAT}) are two of the most important ones for PE Parser. \texttt{Import} points to a squence of structures with each storing information that corresponds to a DLL and the system functions supplied in the DLL when loaded into memory, and is stored in one of the sections. The entry to \texttt{Import} is a relative virtual address (RVA) in the memory. In this study, we should convert all RVAs to file offset addresses (FOAs) in the malware or benignware files because they are not executed. The conversion can be caculated by  
	\begin{equation}\label{RVATOFOA}
	     FOA = RVA - SectionRVA + SectionFOA\cdot
	\end{equation}
	where SectionRVA is the start address of the section in the memory where \texttt{Import} resides, and SectionFOA is the start address of the section in the malware or benignware file. Then, one can locate \texttt{Import}, 
	in which \texttt{Name} is a RVA pointing to the name of a \texttt{DLL}, and \texttt{OriginalFirstThunk} (RVA) points to an import name table (\texttt{INT}) 
	where the function names' addresses are stored. Each address points to a structure called \texttt{\_IMAGE\_IMPORT\_BY\_NAME} at which the function name stays. For the sake of rationality, we filter the function name that is imported as an ordinal when the value of \texttt{Ordinal} is greater than $0x80,000,000$. We then acquire the \texttt{DLL} name and all system function names in the \texttt{DLL} by \texttt{NAME} and \texttt{INT}. \texttt{IAT} has the same functionality as \texttt{OriginalFirstThunk}, and \texttt{FirstThunk} is a pointer to \texttt{IAT}, which can also be searched from \texttt{DataDirectories}. Moreover, each \texttt{Import} or \texttt{IAT} is followed by another \texttt{Import} or \texttt{IAT}. Therefore in order to extract all \texttt{DLL} names and the system function names called by them, PE Parser should repeat the procedure iteratively until no more \texttt{Import} or \texttt{IAT} is available. 
	
	After parsing all malware and benignware, the set of features, i.e., the names of all DLLs and system functions, are collected. Then the feature vector of each program is represented by a binary vector with an entry value '1' indicating that the program possesses the correponding feature (calling the DLL or system function) and '0' otherwise. 
	
	\begin{figure}[!htb]
		\centering
		\includegraphics[width=90mm,height=54mm]{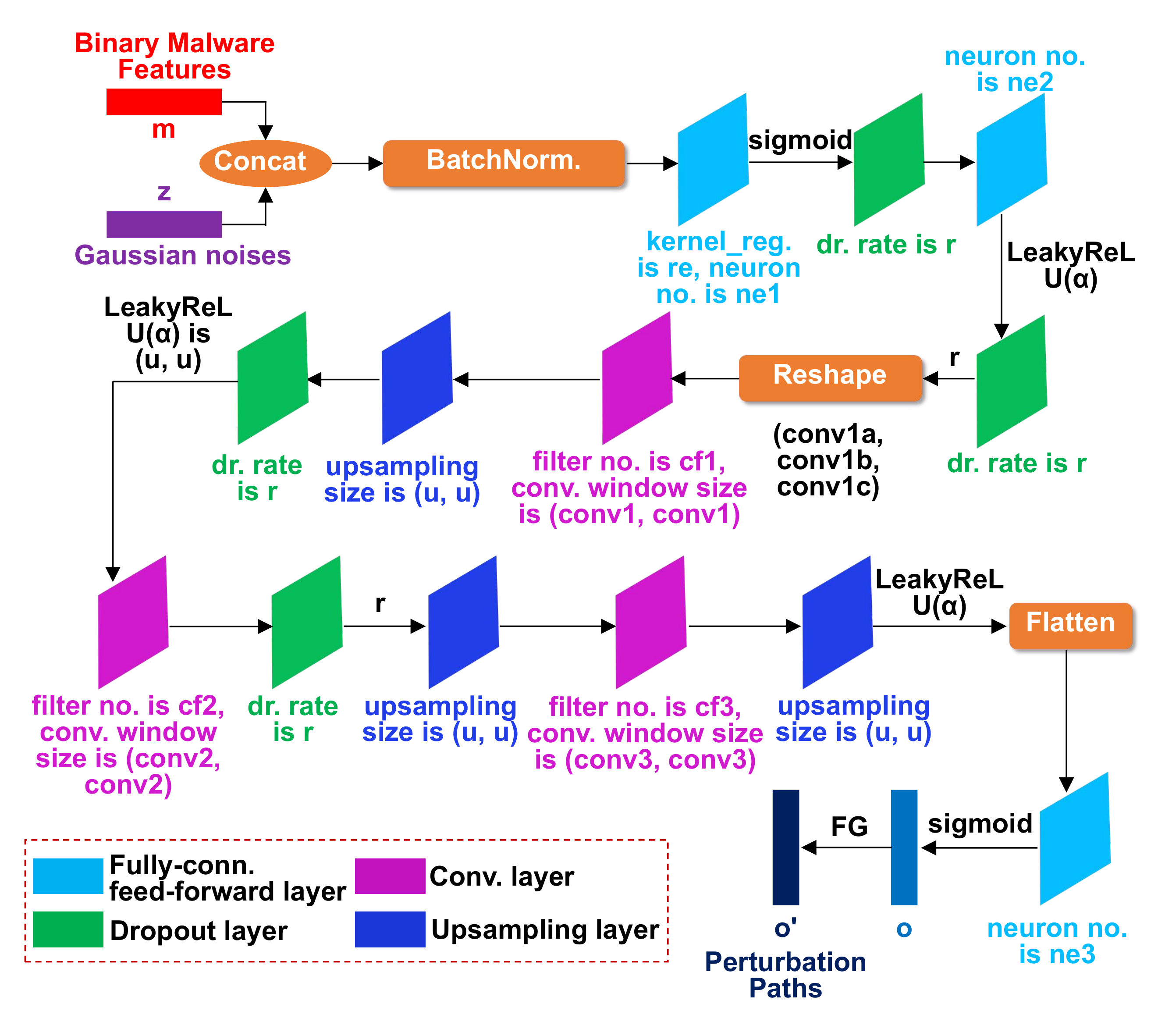}
		\caption{Network Structure of the Generator}
		\label{fig:generator}
	\end{figure}

	\subsection{Generator}\label{sec:Generator}
	
	Generator is used to produce a perturbation path that improves malware's evasive capability while maximizing the probability of Discriminator making a mistake. It employs a transposed CNN with $4$ types of layers: fully connected  feed-forward layers, upsampling layers, dropout layers, and convolutional layers (see Fig. \ref{fig:generator}). As the transposed CNN can utilize its property of composing a specific set of malware features to uniquely represent the malware, Generator makes use of this property to generate the appropriate perturbation path for each malware to help the malware escape from detection. In the following we describe the 13-layer Generator structure employed by our study for the dataset presented in Section~\ref{sec:evaluation}.

	The input to Generator is a malware feature vector concatenated with three random values in the range of $[0, 1)$, followed by a BatchNormalization process. We let each random number correspond to one framework perturbation method, in the order of Obfusmal, Stealmal, and Hollowmal. 
	Each random number is drawn from the Gaussian distribution $\mathcal{N}(0,1)$. Since each framework method has multiple implementations, the value of the corresponding random number specifies the exact instantiated method to be used. For example, if we instantiate 100 Obfusmal methods, then $(0,0.01]$ , $(0.01,0.02]$, $\dots$, $(0.99,1.0)$ correspond to the first, second, $\dots$, the 100th Obfusmal method. 
	
	Next the normalized vector is fed to a fully connected feed-forward layer with $ne1$ neurons, a sigmoid activation function and a l2 regularizer $re$, where $ne1$ is the number of antivirus products and online scan engines VirusTotal has. The output of this layer is sent to the first dropout layer with a drop rate $r$, which is used to ensure that the output is not over-fitting. Following the dropout layer is the second fully connected feed-forward layer with $ne2$ neurons and a LeakyReLU activation function. Note that all the LeakyReLU activation functions used in Generator have the same slope coefficient $\alpha$ and intend to add nonlinearity to Generator. The output of this second fully connected feed-forward layer is the input to the second dropout layer with the same drop rate as the first one. After the second dropout layer, its output is reshaped into a 3-D vector with a shape of $(conv1a, conv1b, conv1c)$ before sent to the first convolutional layer. The first convolutional layer contains $cf1$ filters and a 2D convolution window with a size of $(conv1, conv1)$. Then, the first upsampling layer with a size of $(u, u)$ follows the first convolutional layer. Its output is sent to the third dropout layer with a LeakyReLU activation function, which discards each element with a probability $r$. The output of the third dropout layer passes through the second convolutional layer with $cf2$ filters and a 2D convolution window shaped to $(conv2, conv2)$. The output obtained from the second convolutional layer is sent to the fourth dropout layer with a drop rate $r$. After the dropout layer, the output is upsampled by the second upsampling layer. The third convolutional layer with $cf3$ filters and a 2D convolution window shaped to $(conv3, conv3)$ follows the second upsampling layer before the last upsampling layer with a LeakyReLU activation function.  
	Finally, their output is flattened, and sent to the last fully connected feed-forward layer with $ne3$ neurons and a sigmoid activation function. The sigmoid activation function restricts the values in the generated perturbation path within the range of $(0,1)$. We denote the perturbation path to be $o$. Since the generated perturbation path is supposed to be binary, a smooth function FG is defined to implement binarization tranformation.  The definition of FG is shown in Eq. (\ref{binary}), where each element in $o$ is compared with a threshold 0.5. 
	\begin{equation}\label{binary}
	      FG(o') = o > 0.5.
	\end{equation}
	At last, we obtain a binary vector $o'$  with a shape ($z$, ). Note that $o'$ suggests the customized perturbation path that PE Editor should follow to generate the adversarial malware example for the input malware. For example, if the customized perturbation path $o'$ has a value $(0,0,1)$, PE Editor adopts Hollowmal only (actually the specific Hollowmal method determined by the random number); and if the customized perturbation path $o'$ has a value $(1,0,1)$, it adopts Obfusmal followed by Hollowmal (i.e., the specific Obfusmal method followed by the Hollowmal method determined by the random numbers).
	
	For better elaboration, we denote the data output by the $i$-$th$ layer to be $X^{C}_i$, where $i$ = $1$, $2$, $3$, $\dots$, $13$, the weight vector between the $i$-$th$ layer and the $i$+$1$-$th$ layer to be $W^{C}_{i,i+1}$ and the bias vector of the $i$-$th$ layer to be $B^{C}_i$. Then one can represent the relationship between the data from two neighboring layers in our model as 
	\begin{equation}\label{relu}
	X_{i + 1}^C = \{ _{0.01X_i^C\  otherwise}^{X_i^C \   {\    where\ x > 0} }.
	\end{equation}
	where the relationship is LeakyReLU and $x$ is the element in $X_i^C$. When the relationship between two neighboring layers is sigmoid, it is calculated by 
	\begin{equation}\label{sigmoid}
	X_{i + 1}^C = \frac{e^{X_{i}^C}}{(e^{X_{i}^C}+1)}.
	\end{equation}
	The calulations within fully connected feed-forword layers are performed according to 
	\begin{equation}\label{feedforward}
	X_{i + 1}^C = W^{c}_{i,i+1}{X_{i}^C}+B^{C}_i.
	\end{equation} 
	Suppose at the $i$-$th$ convolutional layer and upsampling layer, we denote the number of filters to be $nf$, the size of the 2D convolution window to be $(f, f)$, and the shape of the upsampling map to be $(u, u)$. Then within convolutional layers and upsampling layers, Eq. (\ref{cnn}) and Eq. (\ref{sampling}) are performed.
	\begin{gather}\label{cnn}
	X_{i}^C = Conv2D(nf, (f, f), X_{i}^C).
	\\
	\label{sampling}
	X_{i}^C = UpSampling2D((u,u), X_{i}^C).
	\end{gather} 
	
	To train Generator, the loss function is defined as shown in Eq. (\ref{generator}), where $S_m$ is a set of malware feature vectors,  $z$ is a three-dimensional vector with each element drawn from a normal distribution $\mathcal{N}(0,1)$, G is Generator, and D is Discriminator. Note that Generator aims at lowering the probability of malware to be detected by Discriminator. To achieve this goal, $L_G$ should be minimized with respect to the weights and parameters of Generator. Thus the gradient information of Generator is updated with respect to the loss value in Discriminator via backpropagation when Discriminator doesn't detect the adversarial malware example. 
	\begin{equation}\label{generator}
	{L_G}{\rm{ = }}{{\rm{E}}_{{\rm{sm}} \in S_m,z \sim {\rm \mathcal{N}}(0,1)}}\log D(G(sm,z)).
	\end{equation}

	\subsection{PE Editor}
	PE Editor is in charge of increasing the evasive ability of malware by following the perturbation path output from Generator to produce an adversarial malware example. Recall that the output $o'$ of Generator has 3 elements, specifying whether each of Obfusmal, Stealmal, and Hollowmal is adopted. The idea of Obfusmal is motivated by application shielding software such as 
	ASPack \cite{ASPACK} and PECompact \cite{PECompact}, which are used to compress and protect programs. We employ encryption instead of compression to futher hide the malware, which can obliviously avoid the detection by statistics-based or  data comparison methods \cite{edgesecurity}. The principles of Stealmal and Hollowmal come from malicious applications NETWIRE Trojan \cite{netwire}, Monero miner \cite{svk}, and BitPaymer Ransomware \cite{bitPaymer}, but we bring encryption to the processes to further conceal the malware. Note that we are not able to find any description on the implementation procedures of the programs mentioned above, and thus largely rely on the public documents to infer their functions. In this aspect, we are the first to reveal the methods. In the following we detail the implementations of these three framework perburbation methods and explain how they are applied. In this paper, we use $malware.exe$ to refer to the malware program that will be converted to a foxy one by MalFox.

	\subsubsection{Obfusmal}
	The implementation of Obfusmal consists of 4 steps: $(i)$ read $malware.exe$, obtain the address and size of its code section, and encrypt the code section; $(ii)$ develop $Shell.dll$ with the functionality that can store the crucial information (code section address, code section size, and decryption key), decrypt the code section, fix up relocation information, and jump to the OEP (an OEP is the start address for program execution) of $malware.exe$ to execute the code; $(iii)$ load $Shell.dll$ into memory to save the address and size of the code section, decryption key, and the OEP of $malware.exe$ in an extern export global variable, and change the OEP of $malware.exe$ to the OEP of $Shell.dll$; $(iv)$ add a section with the length up to $Shell.dll$ in $malware.exe$, save $Shell.dll$ in the newly added section of $malware.exe$, create a buffer with the length up to $malware.exe$ plus $Shell.dll$, copy $malware.exe$ to the buffer and write the buffer into disk. The buffer in the disk contains the foxy malware. The address and size of the code section can be retrieved in the section tables and each section table corresponds to a section in Sections in Fig. \ref{fig:PE}. The section tables follow  Image\_Optional\_Header and each table has the same structure. 
	
	To encrypt the code section, we look for PointerToRawData and SizeOfRawData at the code section table, as therein PointerToRawData stores the file offset address and SizeOfRawData stores the size of the code section. After specifying the address and size of the code section, one can employ any secure encryption algorithm such as DES, AES, 3DES, \emph{etc}. For convinience and simplicity, we utilize the exclusive OR algorithm with the key $0x15$ for encryption. Since the decryption process is taken effect in memory, the address of the encrypted section saved in $Shell.dll$ should be PhysicalAddress. 

	
	
\begin{figure}[!htb]
		\centering
		\includegraphics[width=75mm]{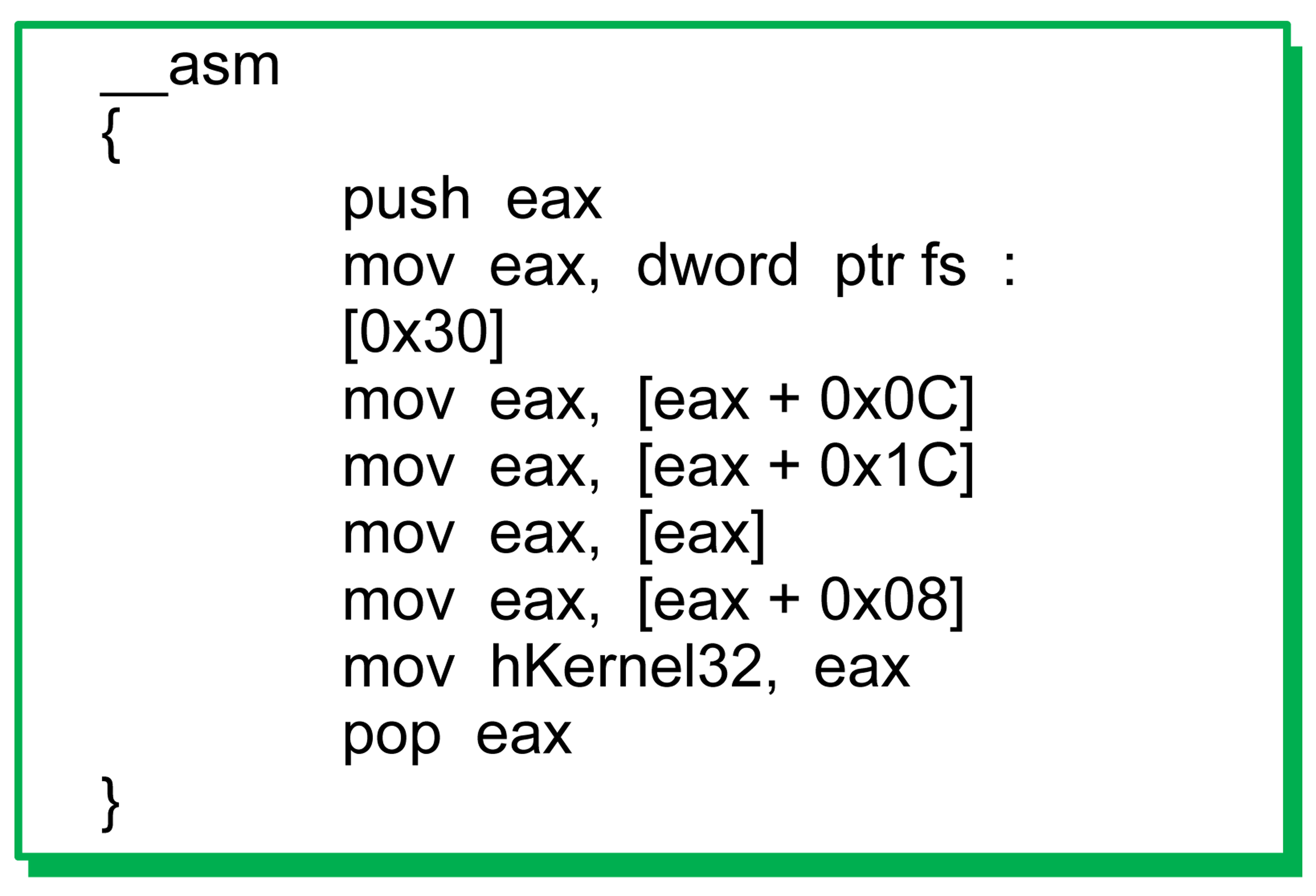}
		\caption{Assembly Code}
		\label{fig:assembly}
	\end{figure}

	When it comes to develop $Shell.dll$, an extern export global variable should be defined to save PhysicalAddress, SizeOfRawData, OEP and the key, such that $Shell.dll$ can decrypt the code section by using the extern export global variable. Next we need to fix up the relocation information since $malware.exe$ might be loaded into an unprefered address that is distinct from the ImageBase field in Image\_Optional\_Header of $malware.exe$, and thus some addresses of the static varibles should be modified. To fix up relocation information, VirtualProtect, VirtualAlloc and GetModuleHandleA supplied by Windows have to be called. However, these functions are no longer provided by our method as the start address of program execution is changed to the OEP of $Shell.dll$ and the execution process of $malware.exe$ loaded into memory through Windows is interrupted. Hence, we have to obtain these functions by resorting to $kernal32.dll$, which is a Windows system kernel DLL that supplies these functions. The base address of  $kernal32.dll$ can be acquired by the assembly code shown in Fig. \ref{fig:assembly}. After $kernal32.dll$ is found, these functions can be retrieved by going through AddressOfName, AddressOfOrdinals and AddressOfFunctions fields in Export\  Tables. 
	Up to now, $Shell.dll$ can use GetModuleHandleA to acquire the start address of $malware.exe$ when loaded into memory. Next, $Shell.dll$ searches for all variable addresses by Reloc\ Tables that need to be fixed up in $malware.exe$. 
	VirtualProtect and VirtualAlloc are used to change the values in variable addresses. Finally, $Shell.dll$ switches the execution power to $malware.exe$. 

	After the development of $Shell.dll$, we read the values of the PhysicalAddress, SizeOfRawData, and OEP fields from $malware.exe$, and load $Shell.dll$ into memory to save these values and the encryption key in the extern export global variable. Next, $malware.exe$ needs to add a section to store $Shell.dll$. The value of the NumberOfSections field in PE\ File\ Header of $malware.exe$ should be increased by 1. Correspondingly the four fields in the last Section Table of $malware.exe$, namely VirtualSize, VirtualAddress, SizeOfRawData, and PointerToRawData, should be set to the value of SizeOfImage in the Image\_Optional\_Header of $Shell.dll$ and that of $malware.exe$, the size of $Shell.dll$ and that of $malware.exe$, respectively. Besides, the execution power should be changed to $Shell.dll$ by revising the OEP of $malware.exe$ to that of $Shell.dll$. In order to change the OEP of $malware.exe$, PE Editor has to retrieve the OEP and SizeOfImage fields in Image\_Optional\_Header of $malware.exe$ and $Shell.dll$. Finally the OEP of $malware.exe$ is changed to the OEP of $Shell.dll$ plus the SizeOfImage of $malware.exe$.
	
\begin{figure}[!htb]
		\centering
		\includegraphics[width=75mm,height=90mm]{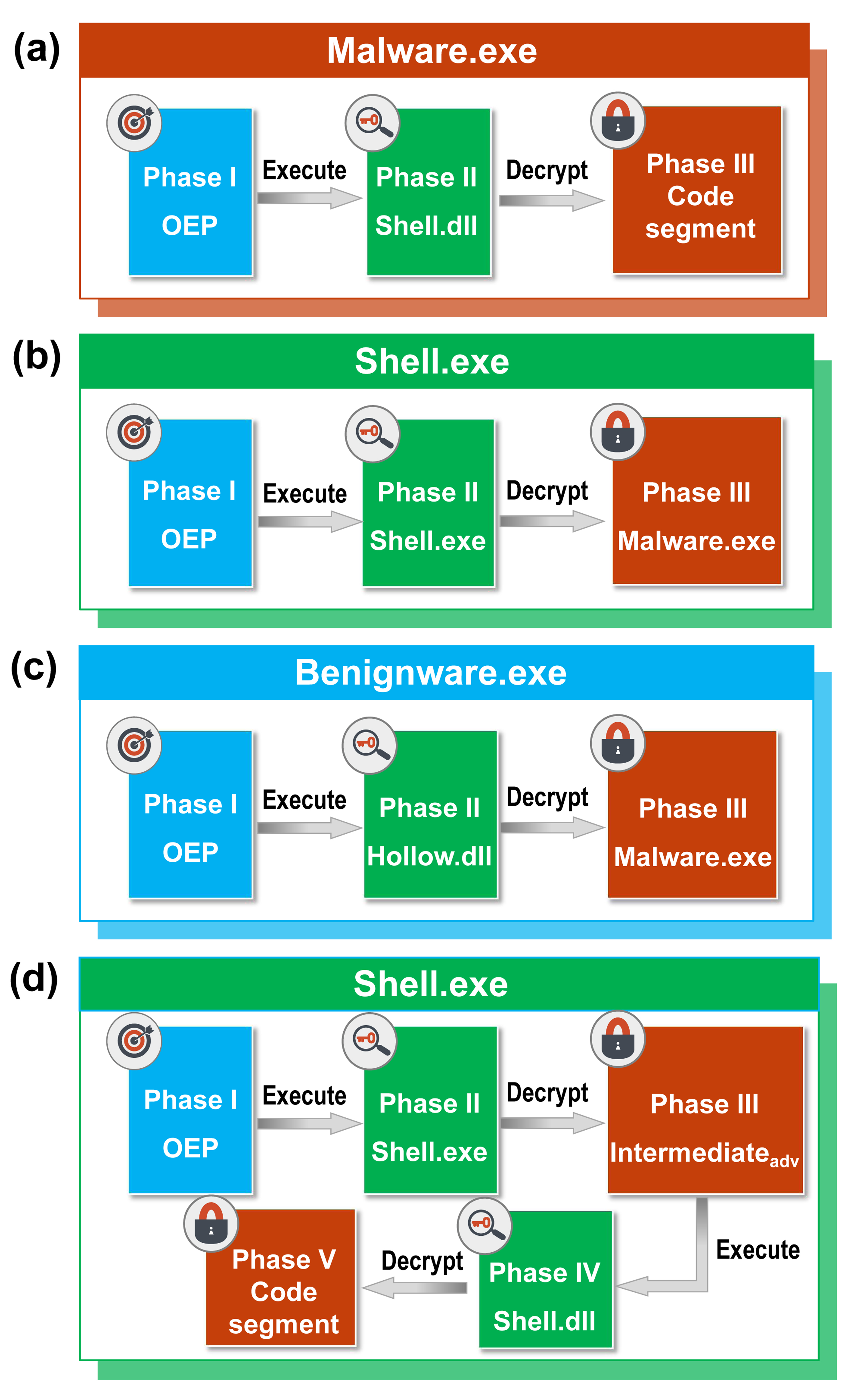}
		\caption{Process of Program Execution}
		\label{fig:PEEDITOR}
	\end{figure}

	Last, PE Parser creates a buffer to save the modified $malware.exe$, writes the buffer into disk, and outputs the foxy malware. If the value of the perturbation path $o'$ is (1, 0, 0), Obfusmal is applied to $malware.exe$ to generate an adversarial malware example. The normal execution flow of the adversarial malware example is illustrated in Fig.\ref{fig:PEEDITOR} (a). In Phase I, the OEP of the adversarial malware example is directed to the start address of $Shell.dll$. In Phase II, $Shell.dll$ decrypts the code section, fixes up the relocation information, and jumps to the OEP of $malware.exe$. In Phase III, the decrypted code section is executed. 
	
	\subsubsection{Stealmal}
	To implement Stealmal, we need to carry out 3 steps: $(i)$ encrypt the entire $malware.exe$; $(ii)$ develop a program $Shell.exe$ and add a section in $Shell.exe$ to save $malware.exe$, where $Shell.exe$ is responsible for decrypting the encrypted $malware.exe$, creating a suspended process, obtaining the process space, copying $malware.exe$ into the space, fixing up the relocation information, and changing the context of the process to the OEP of $malware.exe$; $(iii)$ create a buffer with the length up to $Shell.exe$ plus $malware.exe$ and write the buffer into disk. The last step generates the foxy malware. 
	
	We use the encryption algorithm mentioned above to encrypt $malware.exe$. The encrypted $malware.exe$ is then attached to the end of $Shell.exe$ as a section. Thus, for $Shell.exe$, the value of NumberOfSections field in File\ Header needs to be increased by 1, and PointerToRawData and SizeOfRawData fields in the last Section\ Table should be filled with the size of $Shell.exe$ and that of $malware.exe$, respectively. Besides, VirtualAddress and VirtualSize are respectively assigned values equal to SizeOfImage in Image\_Optional\_Header of $Shell.exe$ and the size of $malware.exe$ accordingly. 
	
	To decipher $malware.exe$, $Shell.exe$ has to search its own last section to locate the encrypted $malware.exe$ with the help of the NumberOfSections field in  File\ Header and the PointerToRawData and SizeOfRawData fields in Section\ Tables. After successfully decrypting $malware.exe$, $Shell.exe$ creates a suspended process by CreateProcess with the sixth parameter valued CREATE\_SUSPENDED, which aims to provide runtime environment for $malware.exe$. To vacate the process space for $malware.exe$, NtUnmapViewOfSection has to be called. However, NtUnmapViewOfSection is a kernel-level application programming interface stored in $ntdll.dll$, thus can't be directly called at the User Layer. To overcome this barrier, we take a tactful strategy by loading $ntdll.dll$ to obtain the function address of NtUnmapViewOfSection, and then defining a function pointer with the same parameters as NtUnmapViewOfSection and referencing it to the function address. Specifically, the parameters are the handle and loading address of the suspended process. The handle is provided by CreateProcess, while the loading address has to be retrieved from Process\ Environment\ Block (PEB). PEB is a user-mode data structure that can be employed by applications to get information such as the list of loaded modules, process startup arguements, heap address, image base address of imported DLLs, etc., \cite{ProcessExecution}. One can acquire the PEB by calling NtQueryInformationProcess, which is also stored in $ntdll.dll$ and can be obtained in the same way as NtUnmapViewOfSection. After the suspended process is uninstalled, $malware.exe$ is copied into the space. Besides, the loading address may be different from the preferred address for $malware.exe$. The relocation information is supposed to be fixed up in the same way as that mentioned in Obfusmal. Next, the context of the suspended process should be changed to the OEP of $malware.exe$. $Shell.exe$ resumes the process, and the power to execute is switched to $malware.exe$. 
	
	The last step operates similarly as Obfusmal, which finally generates the foxy malware. If the value of the perturbation path $o'$ is (0, 1, 0), Stealmal is applied to $malware.exe$ to generate an adversarial malware example, whose execution flow is illustrated in Fig. \ref{fig:PEEDITOR} (b). In Phase I, the OEP of the adversasrial malware example is directed to the start address of $Shell.exe$. In Phase II, $Shell.exe$ decrypts $malware.exe$, creates a suspended process, vacates the process space, copies $malware.exe$ to the space, fixes up the relocation information, and changes the context of the process to the OEP of $malware.exe$. In Phase III, $malware.exe$ is executed.

	\subsubsection{Hollowmal}
	Hollowmal is slightly different from Stealmal. It consists of 3 steps: $(i)$ select benignware, add a section in the benignware with the length up to $malware.exe$, encrypt the entire $malware.exe$, and attach the encrypted $malware.exe$ to the end of the benignware; $(ii)$ develop a DLL named $Hollow.dll$ embracing similar functionality as $Shell.exe$, and add a section in the benginware to store $Hollow.dll$ following $malware.exe$; $(iii)$ create a buffer with the length up to the total size of the benignware, $malware.exe$, and $Hollow.dll$, and write the buffer into disk. 
	
	In our implementation, we choose a very small benignware that only prints out "Hello World", and VirusTotal examine it as benign. To add a section in the benignware so as to store the encrypted $malware.exe$, the values in the NumberOfSections, PointerToRawData, SizeOfRawData,  VirtualAddress, and VirtualSize fields of the last section have to be modified. Then, we encrypt $malware.exe$ employing the same encryption algorithm as mentioned earlier. 
	
	To develop $Hollow.dll$, an extern export global variable is defined to save the encryption key, encryption size, and the address of the benignware to which  $malware.exe$ is attached. The functions of $Shell.exe$ that decrypt $malware.exe$, creat a suspended process, obtain the process space, copy $malware.exe$ into the space, fix up the relocation information, and change the context of the process to the OEP of $malware.exe$ can be moved to $Hollow.dll$, but the section that needs to be decrypted is the second to last section in the benignware. Furthermore, the benignware adds another section attaching $Hollow.dll$ to its end. The values of the NumberOfSections, PointerToRawData, SizeOfRawData,  VirtualAddress, and VirtualSize fields in the benignware are changed correspondingly. 
	
	The last step behaves similarly as that of Obfusmal. Finally it generates the foxy malware. If the value of the perturbation path $o'$ is (0, 0, 1), Hollowmal is applied to $malware.exe$ to generate an adversarial malware example, whose execution flow is illustrated in Fig. \ref{fig:PEEDITOR} (c). In Phase I, the OEP of the adversarial malware example is directed to the start address of $Hollow.dll$. In Phase II, $Hollow.dll$ decrypts $malware.exe$, creates a suspended process, vacates the process space, copies $malware.exe$ to the space, fixes up the relocation information, and changes the context of the process to the OEP of $malware.exe$. In Phase III, $malware.exe$ is executed. 
	
	\subsubsection{Combination}
	When the output of Generator is a path that involves more than one  perturbation method, PE Editor needs to combine and execute all of them when generating the adversarial malware example for $malware.exe$. For example, when the value $o'=(1, 1, 0)$, PE Parser processes $malware.exe$ by employing Obfusmal to generate an intermediate version of the adversarial malware example $Intermediat{e_{Adv}}$, which should be the input to generate the final version of the adversarial malware example $Fina{l_{Adv}}$ by Stealmal. The execution flow of $Fina{l_{Adv}}$ is illustrated in Fig. \ref{fig:PEEDITOR} (d). In Phase I, the $OEP$ of $Fina{l_{Adv}}$ is directedd to the start address of $Shell.exe$. In Phase II, $Shell.exe$ decrypts $Intermediat{e_{Adv}}$, creates a suspended process, vacates the process space, copies $Intermediat{e_{Adv}}$ to the space, fixes up the relocation information, and changes the context of the process to the OEP of $Intermediat{e_{Adv}}$. In Phase III, the OEP of $Intermediat{e_{Adv}}$ is directed to the start address of $Shell.dll$. In Phase IV, $Shell.dll$ decrypts the code section, fixes up the relocation information, and jumps to the OEP of $malware.exe$. In Phase V, the decrypted code section is executed. 
	
	\begin{figure}[!htb]
		\centering
		\includegraphics[width=90mm,height=54mm]{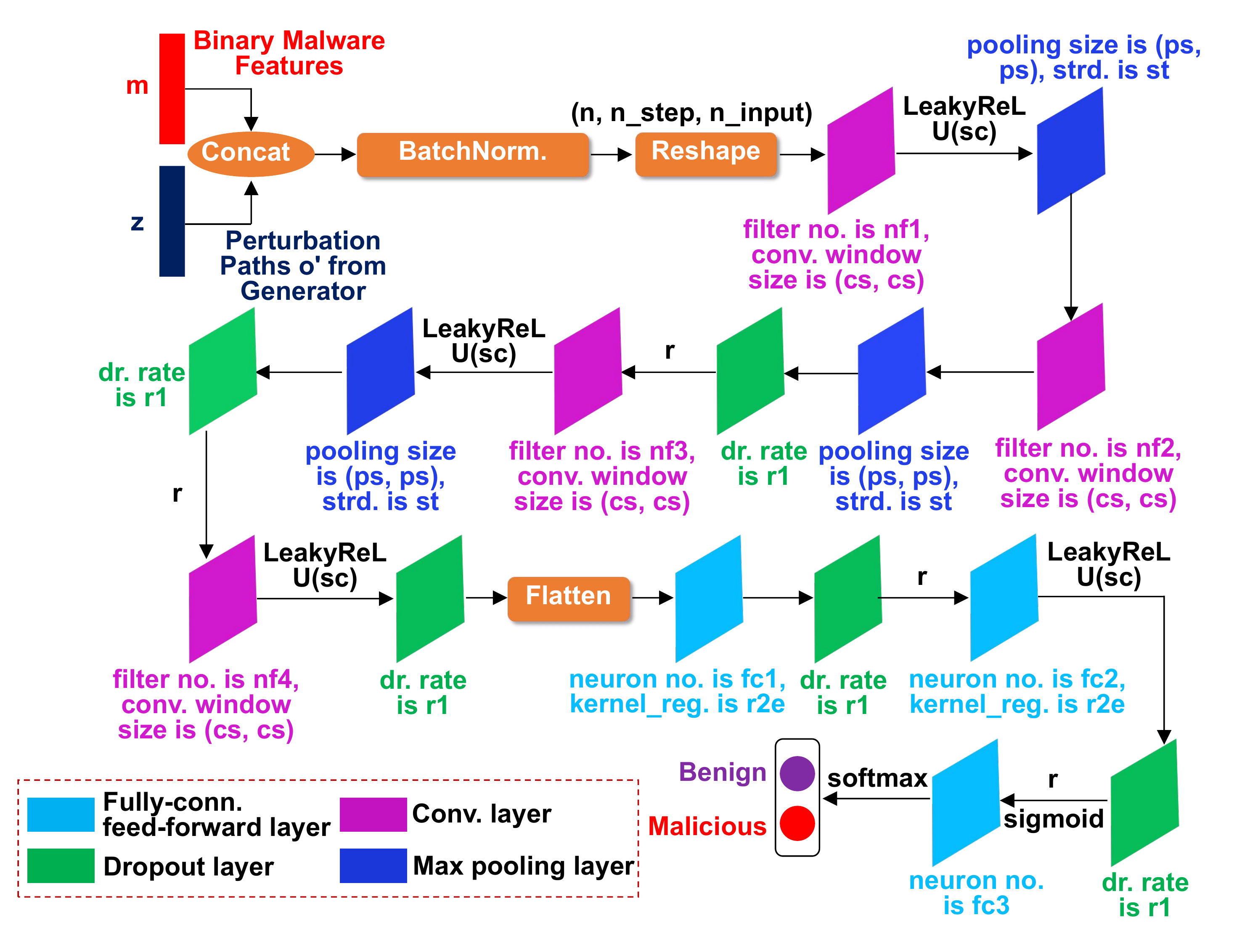}
		\caption{Network Structure of Discriminator}
		\label{fig:discriminator}
	\end{figure}
	
	\subsection{Discriminator}
	Discriminator estimates the probability that a malware program is recognized as benign, and provides gradient information to train Generator. Discriminator employs a CNN algorithm with  similar types of layers as Generator except that the upsampling layers are substituted by max-pooling layers. The structure of Discriminator employed by our study for the dataset presented in Section~\ref{sec:evaluation} is shown in Fig. \ref{fig:discriminator}, which contains 15 layers. CNN is adopted here because it has demonstrated promising feature extraction capability in many fields such as snapshot compressive imaging \cite{sci}, which could precisely distinguish benignware and malware for our purpose.

	Discriminator is constructed with a training dataset that contains both benignware and malware. The input is a 1-D vector obtained by concatinating the binary feature vectors of all training programs. This vector is first normalized via BatchNormalization, then resized into a 3-D vector with a shape of $(n, n\_step, n\_input)$, where $n$ is the number of samples being fed into the network, and the other two parameters are used to meet the shape requirements of the first convolutional layer. The first convolutional layer contains $cf1$ zero padded filters and a convolution window with a shape of $(conv1, conv1)$. The activation function used here is LeakyReLU with a slope coefficient $sc$. The following LeakyReLU activation functions use the same setting. One can get the output with one more dimension in depth, \emph{i.e.}, $X_{1}^C$ has the shape of $(n, n\_step, n\_input, nf1)$. Following the first convolutional layer is the first max-pooling layer with a stride $st$ and a pooling size $(ps, ps)$. After the first max-pooling layer, we have two sets of convolutional layers, max-pooling layers and dropout layers. The two max-pooling layers have the same settings as the first max-pooling layer, while the second and the third convolutional layers are different in the number of filters, \emph{i.e.}, $cf2$, $cf3$, and the third convolutional layer consists of a LeakyReLU activation function. Besides, all dropout layers used in Discriminator have the same drop rate $r$. The output of the second dropout layer is an input to the fourth convolutional layer with $cf4$ filters, a convolution window shaped to $(conv1, conv1)$, and a LeakyReLU activation, before it is sent to the third dropout layer. The output of the third dropout layer is flattened, and then sent to the first fully connected feed-forward layer with $fc1$ neurons and a l2 regularizer $r2e$. Following the first fully connected feed-forward layer is the fourth dropout layer and the second fully connected feed-forward layer consisting of $fc2$ neurons, a l2 regularizer $re$, and a LeakyReLU activation function where $fc2$ is the number of antivirus products and online scan engines VirusTotal has. The last dropout layer contains a sigmoid activation function following the second fully connected feed-forward layer. The output of the last dropout layer is fed into the last fully connected feed-forward layer consisting of a softmax activation function (see Eq. (\ref{softmax})), which finally outputs the features with a shape ( $fc3$, ), where $fc3$ corresponds to the classification of the programs. 
	\begin{equation}\label{softmax}
	X_{{i_{j'}}}^C =\frac {\exp \left( {X_{{{\rm{i}}_j}}^C} \right)}{{\Sigma _k}\exp \left( {X_{{{\rm{i}}_k}}^C} \right)}.
	\end{equation}
	
	The loss function of Discriminator is  defined in Eq. (\ref{discriminator}), where $S_b$ is a set of benignware feature vectors, and $S_m$ is a set of malware feature vectors.
	\begin{equation}\label{discriminator}
	{L_D}{\rm{ =  - }}{{\rm{E}}_{{\rm{sb}} \in S_b}}\log (1 - D(sb)) - {E_{sm \in S_m}}\log (D(sm)).
	\end{equation}
	To precisely distinguish benignware and malware in the dataset, $L_D$ should be minimized with regard to the weights and parameters of Discriminator. Currently, one can see that minimizing $L_G$ would lower the predicted malicious probability of malware, and force Discriminator to recognize malware as benign. In our design, Discriminator adjusts itself to have the same detective capability as VirusTotal to differentiate benignware and malware. As a result, the continuous training of Generator to generate adversarial malware examples needs to first confuse Discriminator and then further fool VirusTotal. 
	
	When employing Discriminator to test an adversarial malware example, the input is the concatination of the malware feature vector and the corresponding perturbation path for that malware, and the output is either benign or malicious indicating the produced adversarial malware example is incorrectly or correctly detected.

	\section{Evaluation}
	\label{sec:evaluation}
	The metrics to measure the performance of MalFox are \emph{accuracy}, \emph{detection rate} and \emph{evasive rate} via Discriminator or VirusTotal. Accuracy is the ratio of incorrectly predicted adversarial malware examples ($a$) over all adversarial malware examples ($A$) by Discriminator (see Eq. (\ref{accuracy})). Detection rate is the ratio of entities ($n$) that detect the malware or adversarial malware example over all entities ($N$) in VirusTotal (see Eq. (\ref{detectionrate})). Evasive rate is computed by Eq. (\ref{evasiverate}), where ${N_{orig}}$ is the number of entities that detect the malware, and ${N_{adv}}$ is the number of entities that detect the corresponding adversarial malware example. The higher the evasive rate, the higher the probability of the adversarial example being recognized as benign.
	\begin{equation}\label{accuracy}
	accuracy = \frac{a}{A}.
	\end{equation}
	\begin{equation}\label{detectionrate}
	detection\ rate = \frac{n}{N}.
	\end{equation}
	\begin{equation}\label{evasiverate}
	evasive\ rate = \frac{N_{orig}-N_{adv}}{N_{orig}}.
	\end{equation}

	\subsection{Experimental Setup}

	\textbf{Datasets.} In our experiment, we collected a malicious dataset of 13425 malware from VirusShare \cite{VirusShare} and a benign dataset of 1719 benignware from Only Freeware \cite{freeware}, SnapFiles \cite{SnapFiles}, and downloadcrew \cite{downloadcrew}. As shown in Table \ref{tbl:tab1}, the test dataset of the Generator includes 1000 malware that are randomly chosen from the malicious dataset while the training dataset includes all the benignware as well as the leftover 12425 malware. 
	\begin{table}[!htb]
	\captionsetup[table]{skip=10pt}
	\caption{The Training and Test Datasets for the Generator}\label{tbl:tab1}
		\centering
		\begin{tabular}{|c|c|c|}
			\hline
			--          & Malicious Dataset & Benign Dataset \\ \hline
			Training Dataset & 12,425     & 1,719     \\ 
			\hline
			Test Dataset  & 1,000    & 0     \\ 
			\hline
		\end{tabular}
	\end{table}
	
	\begin{table}[!htb]
	\captionsetup[table]{skip=10pt}
	\caption{Options of the Perturbation Path}\label{tbl:tab6}
		\centering
		\begin{tabular}{|c|}
			\hline
			perturbation path           \\ \hline
			(0, 0, 0)        \\ 
			\hline
			(1, 0, 0)         \\ 
			\hline
			(0, 1, 0)         \\ 
			\hline
			(0, 0, 1)         \\ 
			\hline
			(1, 1, 0)         \\ 
			\hline
			(1, 0, 1)         \\ 
			\hline
			(0, 1, 1)         \\ 
			\hline
			(1, 1, 1)         \\ 
			\hline
		\end{tabular}
	\end{table}
	
	Both malware and benignware need to be screened to ensure the relability of our dataset. For this purpose we employ VirusTotal to test all of them, and regard a program as malware if it is detected to be milicous by one of the 82 entities (antivirus software and online scan engies)\footnote{VirusTotal may contain more than 82 entities but the results of our submissions indicate that 82 entities examined our programs.} in VirusTotal while it is benign otherwise. Additionally PE Parser makes use of 8 options (see Table \ref{tbl:tab6}) of the perturbation path $o'$ according to the order of Obfusmal, Stealmal, and Hollowmal discribed in Section~\ref{sec:PE:Parser}. For example, when $o'$ is (1,0,1), $malware.exe$ is first processed by Obfusmal and then by Hollowmal while Stealmal is ignored, and when $o'$ is (1,1,1), $malware.exe$ is processed by all three perturbation schemes in the order of Obfusmal, Stealmal, and Hollowmal. Note that in this experimental study, we instantiate each framework method with only one implementation for demonstration purpose only. 
	
	Codes are available upon request -- we are happy to share the codes with researchers who are interested in relevant research. Written in C++, PE Parser consists of 541 lines of code, and PE Editor consists of 2565 lines of code; written in Python, Generator and Discriminator contains 850 lines of code in total. 
	

We conduct 30 independent simulation trials for statistical confidence. The training process follows Algorithm~\ref{malfox}. The accuracies of MalFox during the training and testing processes are evaluated via Discriminator. For the test dataset, the detection rate and evasive rate of each malware and its corresponding adversarial malware example are computed by collecting the detection results after uploading it to VirusTotal. At last, the averaged detection rate and evasive rate are calculated.

	\textbf{Model Parameters.} In order to identify a proper structure to realize our goals, we attempt the number of layers for both Generator and Discriminator from $1$ to $20$. It turns out that a structure with $13$ layers for Generator and one with $15$ layers for Discriminator are the most effective ones in terms of accuracy. Specifically, Generator includes $3$ fully connected feed-forward layers, $4$ dropout layers, $3$ upsampling layers, and $3$ convolutional layers; and Discriminator includes $4$ convolutional layers, $3$ max-pooling layers, $5$ dropout layers, and $3$ fully connected feed-forward layers. Model parameter details for Generator and Discriminator are presented in Table \ref{tbl:tab2} and Table \ref{tbl:tab3}, respectively.
	
	\begin{table}[!htb]
		\captionsetup[table]{skip=10pt}
		\caption{Model Parameters for Generator}\label{tbl:tab2}
	\centering    
	\begin{tabular}{|m{2cm}<{\centering}|m{3cm}<{\centering}|m{2cm}<{\centering}|}
		\hline
		Parameters & Implications  & Values \\ 
		\hline
		$m$    & num of features & 16156\\ 
		\hline
		$|z|$     & num of perturbation methods  & 3 \\ 
		\hline
		$ne1$    & num of entities in VirusTotal & 82\\ 
		\hline
		$re$     &   penalty  & 0.01 \\ 
		\hline
		$r$     &  drop rate  & 0.5 \\ 
		\hline
		$ne2$   & num of neurons  & 1248 \\ 
		\hline
		$\alpha$  & slope coefficient  & 0.1 \\ 
		\hline
		$conv1a$   & shape size  & 8 \\
		\hline
		$conv1b$   & shape size  & 13 \\
		\hline
		$conv1c$   & shape size & 12 \\
		\hline
		$cf1$   & num of filters   & 32 \\
		\hline
		$conv1$   & shape  & 5 \\
		\hline
		$u$   & repeatitions  & 2 \\
		\hline
		$cf2$   & num of filters & 64 \\
		\hline
		$conv2$   &  shape  & 2 \\
		\hline
		$cf3$   & n um of filters & 256 \\
		\hline
		$conv3$   & shape & 2 \\
		\hline
		$ne3$   &  dimension  & 3 \\
		\hline
	\end{tabular}
	
\end{table}

\begin{table}[ht]
	\captionsetup[table]{skip=10pt}
	\caption{Model Parameters for Discriminator}\label{tbl:tab3}
	\centering    
	\begin{tabular}{|m{2cm}<{\centering}|m{3cm}<{\centering}|m{2cm}<{\centering}|}
		\hline
		Parameters & Implications  & Values \\ 
		\hline
		$n$     & num of samples  & 32 \\ 
		\hline
		$n\_step$  &  shape   & 143 \\ 
		\hline
		$n\_input$  & shape  & 113 \\
		\hline
		$cf1$   & num of filters  & 512 \\
		\hline
		$conv1$   & convolution window  & 2 \\
		\hline
		$sc$   & slope coefficient  & 0.1 \\
		\hline
		$st$   & num of pooling window moves  & 1 \\
		\hline
		$ps$   & pooling size  & 2 \\
		\hline
		$cf2$   & num of filters & 256 \\
		\hline
		$r$   & drop rate & 0.5 \\
		\hline
		$cf3$   & num of filters   & 64 \\	
		\hline
		$cf4$   & num of filters  & 32 \\
		\hline
		$fc1$   & num of neurons & 1024 \\
		\hline
		$re$   & penalty  & 0.01 \\
		\hline
		$fc2$   & num of entities & 2 \\
		\hline
		$fc3$   & num of classes & 2 \\
		\hline
	\end{tabular}
	
\end{table}

	\subsection{Evaluation Results}
	\subsubsection{Accuracy}
	\label{sec:accuracy}
	\begin{figure}[!htb]
		\centering
		\includegraphics[scale=0.51]{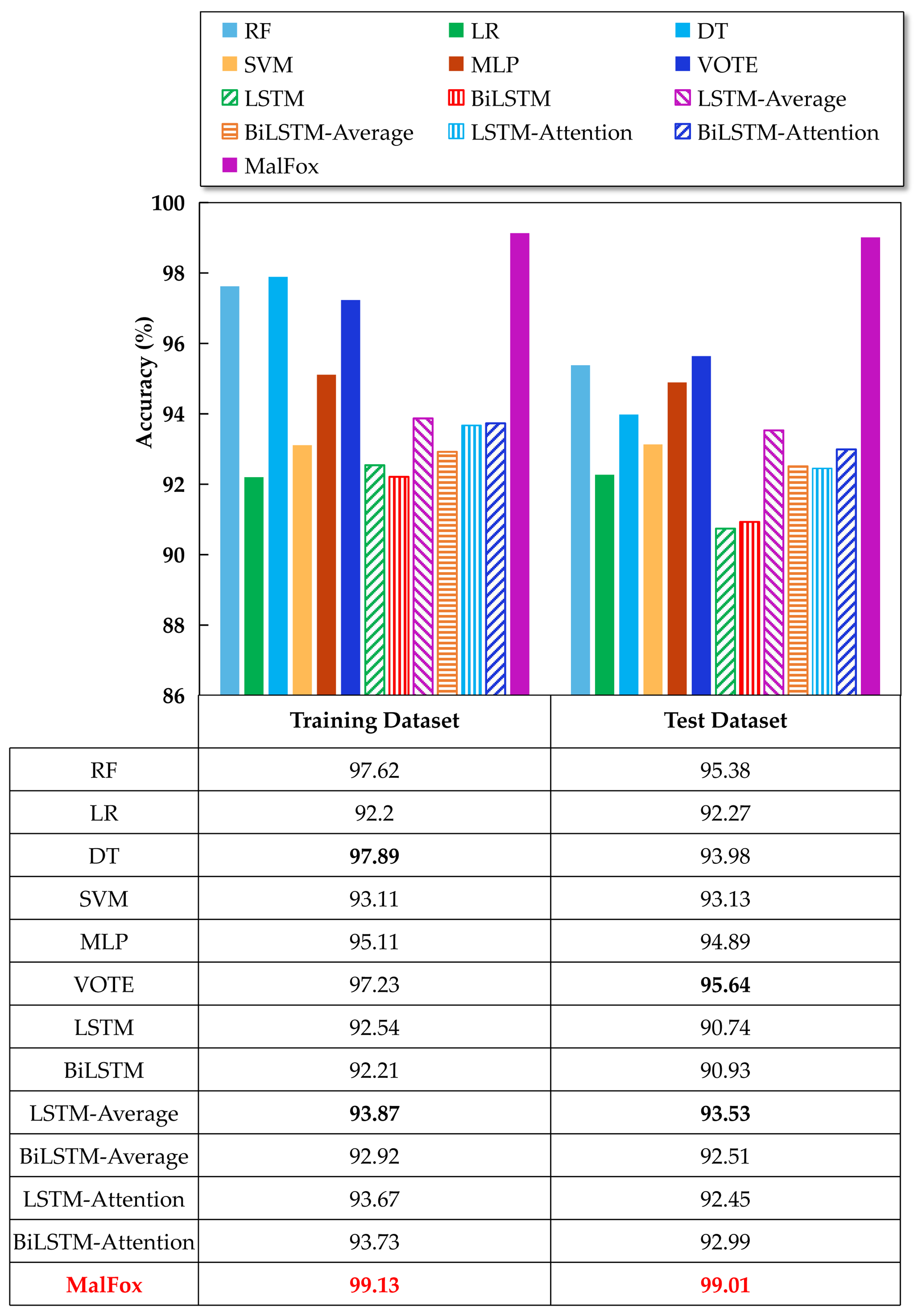}
		\caption{Evaluation Results on Accuracy}
		\label{fig:accuracy}
	\end{figure}
	
	We first evaluate the accuracy of MalFox with the collected dataset that contains in total 15144 programs. When taking Discriminator as the attack target, we compare the accuracy of MalFox against 6 machine learning models (Random Forest (RF), Logistic Regression (LR), Decision Tree (DT), Support Vector Machine (SVM), Multi-layer Perceptron (MLP), and a voting based ensemble of these classifiers (VOTE)  \cite{malgan}), and 6 deep learning models (LSTM, BiLSTM, LSTM-Average, BiLSTM-Average, LSTM-Attention, and BiLSTM-Attention \cite{lstmgan}) that also employ generative adversarial networks to produce adversarial malware examples. The results indicate that our MalFox framework has an average accuracy of 99.13\% for the training dataset and 99.01\% for the test dataset, which outperform the contrastive machine learning models whose highest accuracy for training is 97.89\% and for testing is 95.64\%, and the contrastive deep learning models whose highest accuracy for training is 93.87\% and for testing is 93.53\%. Detailed comparison results are shown in Fig. \ref{fig:accuracy}.

	\subsubsection{Detection Rate And Evasive Rate}
	\label{sec:deteva}
	
	In this subsection, we evaluate the performance of MalFox via detection rate and evasive rate. Low detection rate and high evasive rate indicate the effectiveness of adversarial malware examples to avoid detection by VirusTotal. Since each user normally installs one or two malware detectors on its devices, our attacks are considered to be effective if the adversarial malware example causes lower detection rate than the malware, and has positive evasive rate. Therefore it is reasonable for us to use the detection rate and evasive rate to evaluate the performance of MalFox. \par
\begin{figure}[!htb]
		\centering
		\includegraphics[width=75mm,height=41mm]{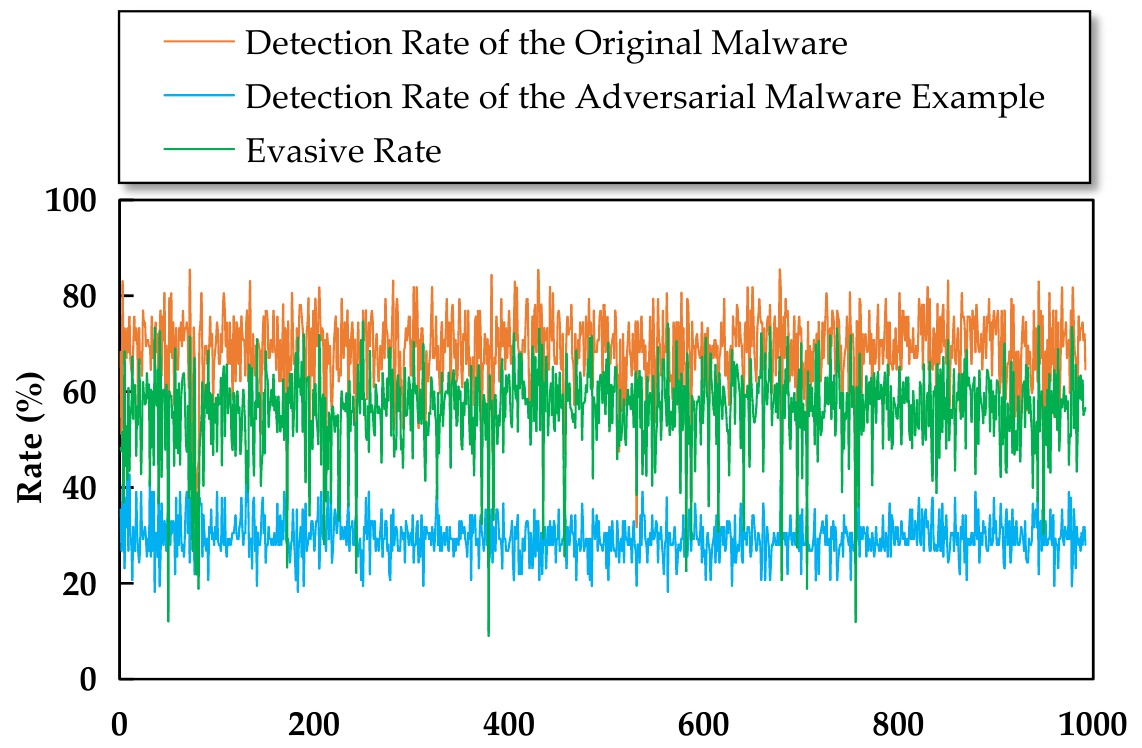}
		\caption{Evasive Rate and Detection Rate}
		\label{fig:evasive}
	\end{figure}

\begin{table}[!htb]
	\captionsetup[table]{skip=10pt}
	\caption{Comparison Results}\label{tbl:tab5}
	\centering
	\begin{tabular}{|c|c|c|c|}
		\hline
		Evaluation Metrics                            & Average & Max & Min\\ 
		\hline
		Detection Rate (Malware)             & 68.8 & 85.4  &  26.8     \\ 
		\hline
		Detection Rate (Foxy Malware)      & 29.7 & 43.9 & 18.3   \\ 
		\hline
		Evasive Rate  (Foxy Malware)       & 56.2 & 74.6 & 9.1\\
		\hline
	\end{tabular}	
\end{table}
Anti-virus products and scan engines in VirusTotal were developed by different companies or institutes, and they examine malware based on different techniques. Therefore for single malware, it might be detected by some entities, but escape detection from others. The detection rate cannot  reach 100\% for the malware. When comparing malware and their adversarial malware examples in Table \ref{tbl:tab5}, one can see that the average detection rate for malware has dropped dramatically by about 56.8\% after being processed by $\MalFox$. Correspondingly, the average evasive rate for adversarial malware examples has been significantly improved (about 56.2\%). Specifically, as suggested in Fig. \ref{fig:evasive}, the detection rate for malware ranges from 26.8\% to 85.4\%. However, the detection rate of their adversarial malware examples are far lower than the original malware, which is between 18.3\% and 43.9\%. Speaking of single malware, the detection rate for its corresponding adversarial malware example has a significant decline in most cases, and the largest drop even reaches 61.0\%. Moreover, the highest evasive rate is 74.6\%, which means that 74.6\% of the detection entities for the malware report the corresponding adversarial malware example as benign, though they previously label the original malware malicious.
	
	\section{Conclusion and Future Research}
	\label{sec:conclusion}
	In this paper, we propose a novel Conv-GAN-based adversarial malware example generation framework titled MalFox, which can transform a malware program into a foxy one with a significantly higher chance of evading detection. MalFox is flexible to accept any number and type of perturbation methods, thus enlarging the options for more effective perturbation paths to enhance malware ability of escaping detection. On the other hand, MalFox successfully attacks practical malware detectors without knowing their underlying implementation details, which renders it a more practical and powerful attack framework.

	In our future research, we will consider more perturbation methods as well as  other powerful evasion tricks such as junk instructions and anti-debugging \cite{debugging}, which can be added into adversarial malware examples to improve their evasive abilities. We also intend to conduct fuzzing tests \cite{fuzzing} with adversarial malware examples generated by different perturbation paths to expose weaknesses of commercial anti-virus products or scan engines, which would contribute to enhance their reliability.
	
%
	
	\ifCLASSOPTIONcompsoc

\bibliographystyle{IEEEtran}
\bibliography{paper}

\end{document}